\documentclass{ws-ijmpe}
\usepackage{graphicx}
\usepackage{epsfig}
\usepackage{psfrag}
\begin{document}

\markboth{M. M. Saez, O. Civitarese, M. E. Mosquera}{
Neutrino mixing in nuclear rapid neutron-capture processes}

\catchline{}{}{}{}{}

\title{Neutrino mixing in nuclear rapid neutron-capture processes}

\author{M. M. Saez}

\address{Facultad de Ciencias Astron\'omicas y Geof\'{\i}sicas, University of La Plata. Paseo del Bosque S/N\\
1900, La Plata, Argentina.\\
msaez@fcaglp.unlp.edu.ar}

\author{O. Civitarese}

\address{Dept. of Physics, University of La Plata, c.c.~67\\
 1900, La Plata, Argentina\\
osvaldo.civitarese@fisica.unlp.edu.ar}

\author{M. E. Mosquera}

\address{Dept. of Physics, University of La Plata, c.c.~67\\
Facultad de Ciencias Astron\'omicas y Geof\'{\i}sicas, University of La Plata. Paseo del Bosque S/N\\
 1900, La Plata, Argentina\\
mmosquera@fcaglp.unlp.edu.ar
}
\maketitle

\begin{history}
\received{Day Month Year}
\revised{Day Month Year}
\end{history}

\begin{abstract}
A possible mechanism for the formation of heavy-mass elements in supernovae is the rapid neutron-capture-mechanism (r-process).
It depends upon the electron-fraction $Y_e$, a quantity which is determined by beta-decay-rates.
In this paper we focus on the calculation of electroweak decay-rates in presence of massive neutrinos. The resulting expressions are then used to calculate nuclear reactions entering the rapid-neutron capture. We fix the astrophysical parameters to the case of a core-collapse supernova. The neutrino sector includes a mass scheme and mixing angles for active neutrinos, and  also by including the mixing between active and sterile neutrinos.
The results of the calculations show that the predicted abundances of heavy-mass nuclei are indeed affected by the neutrino mixing.
\end{abstract}
 \ccode{ 26.30.-k,26.30.Hj,26.50.+x,26.30.Jk}

\maketitle

\section{Introduction}
\label{Intro}

Neutrinos are involved in the chain of reactions leading to the production of light-nuclei, i.e; during the Big Bang Nucleosynthesis, as well as in the production of heavier elements in explosive astrophysical environments \cite{wu:2015,qian:1993,woosley:1994}. Neutrinos of different flavors interact with matter and any mechanism which modifies their composition can potentially affect the nuclear abundances. Since neutrinos oscillates between mass-eigenstates the rates of reactions of cosmological and astrophysical interest depend on oscillation parameters \cite{civitarese:2014,mosquera:2011,mosquera:2015,saez:2018}.

Experiments with solar, atmospheric and reactor neutrinos have provided evidence of the oscillation between neutrino flavors caused by non-zero neutrino masses. The oscillations of neutrinos have been observed by different collaborations such as LSND (Liquid Scintillator Neutrino Detector), SK (SuperKamiokande), SNO (Sudbury Neutrino Observatory), among others \cite{aguilar:2001,fukuda:1998,poon:2001,aguilar:2007,aguilar:2013,palomares:2009,abdurashitov:2009,detwiler:2003,arpesella:2008}.

The results published by LSND and MiniBoone (Mini Booster Neutrino Experiment) collaborations have established limits for the existence of another extra type of neutrino, called the sterile neutrino \cite{eitel:2000,athana:1995,athanassopoulos:1996,aguilar:2018}. The data suggest a mass-square difference between the
lightest mass eigenstate ($m_1$) and the sterile neutrino ($m_4$) of the order
of $\Delta m_{14}^2 \geq 1.5\, {\rm eV}^2$ and a mixing angle  $ \sin^2\theta_{14}\sim 0.14 $ \cite{giunti:2011,himmel:2015}.

The consequences of the existence of sterile neutrinos in different astrophysical scenarios have being examined in previous works by  Boyarsky \cite{boyarsky:2009}, Mohapatra \cite{mohapatra:2004}, and Raffelt \cite{raffelt:1987}, among other authors. In particular, in the context of supernovae (SN), flavor neutrinos may convert into sterile ones in regions near the core of the SN. It is believed that this can lead to a decrease of the electron-neutrino fluxes \cite{molinari:2003}. The effects of oscillations between  active neutrinos and between active and sterile ones in SN explosions and during the neutron-cooling phase of the proto-neutron-star, and the impact of active-sterile oscillations in the development of a successful r-process in the outflow of neutrinos, or neutrino wind, have been studied by several authors \cite{wu:2015,woosley:1994,balasi:2015,fetter:2003,balantekin:2004,tamborra:2012,janka:2012,pastor:2002,qian:2003}.

The abundances of elements in regions near A = 130, and in the area of the actinides, are consistent with the process of rapid neutron capture \cite{anders:1989}. There exist several astrophysical systems that could produce these elements, such as: neutron stars and neutron star fusions, mergers of black-holes and neutron stars and supernova events by nuclear collapse \cite{duan:2010}. Ejecta from neutron star mergers become of special interest after the recent detection of a neutron star merger, however, a weak r-process can occur in a complementary way in core-collapse SN \cite{curtis:2018,cowan:2019}.

In this work, we focus our analysis on the r-process which produces
unstable nuclei that rapidly decay through a series of beta decays
until they become stable. As said before, this process is relevant
in stellar nucleo-synthesis and SN events \cite{clayton:1968}. For
the r-process to take place, it is necessary to have a powerful
current of neutrons in a high temperature environment. Due to the
high neutron-flux, the rate of isotopic formation is greater than
the rate of the subsequent beta decay, therefore the elements
created by this path ascend rapidly through the $N/Z$ stability
line.

This work is organized as follow. In Section \ref{formalism} we
present a brief description of the formalism needed to compute the
beta-decay rates in presence of massive neutrinos. In Section
\ref{resultados} we describe the details of the calculation of the
abundances  of heavy nuclei in a core-collapse supernova
environment. Finally, the conclusions are drawn in Section
\ref{conclusiones}.

\section{Neutrino oscillations and $\beta$-decay rates}
\label{formalism}

We start by written the Hamiltonian density \cite{marshak:1969,blin-stoyle:1973}
\begin{eqnarray}\label{hamil}
\mathcal{H}_\beta=\frac{G_F}{\sqrt{2}}V_{ud}\left(J_\mu L^{\dagger^\mu}+hc \right)\, \, \, ,
\end{eqnarray}
where $G_F$ is the Fermi coupling constant, $V_{ud}=\cos(\theta_{Cabibbo})=0.9738\pm 0.0005$ \cite{yao:2006}. $J_\mu$ and $L_\mu$ are the hadronic and leptonic currents, respectively.

The transition amplitude is written
\begin{eqnarray}\label{abeta}
A_{\beta}(n\rightarrow p+e^-+\bar{\nu_e})&=&< p \,\, e^-\,\, \bar{\nu_e}| \int{d^4x}\mathcal{H_{\beta}} |n > \, \, \, .
\end{eqnarray}
In order to obtain the beta-decay-rate we need to computed  $\left|A_{\beta}\right|^2$.
Since the electron-neutrino is a composite particle we write $|\nu_e>=\sum_j U_ {ej} |\nu_j>$, where $U_{ij}$ is the $3 \times 3$ mixing matrix between active-neutrino mass eigenstates and where we have set Dirac's and Majorana CP-violating phases at zero.
Replacing it in Eq.(\ref{abeta}) leads to
\begin{eqnarray}\label{abetaj}
A_{\beta}\left(n\rightarrow p+e^-+\bar{\nu_e}\right)&=&\sum_j U_ {ej} A_j\left(n\rightarrow p+e^-+\bar{\nu_j}\right)\, \, \, ,
\end{eqnarray}
as done in \cite{ivanov:2008}. Furthermore, and in order to perform the calculations, we have expressed neutrino-mass eigenstates
following  Ref. \cite{ivanov:2008} and considered a Gaussian package with a radial spreading $\delta$
\begin{eqnarray}\label{ivanov1}
<\bar{\nu_j}|{\psi}_{\nu_j}(x,t)&=&<0|\sqrt{\frac{1}{2 q_{\nu_j}^0}} \int\frac{d^3q_{\nu_e}}{(2\pi)^3} e^{-(\bar{q}_{\nu_e}-\bar{q}_{\nu_j})^2\frac{\delta^2}{2}} e^{-\imath \bar{q}_{\nu_e} \cdot \bar{r}+ \imath q_{\nu_j}^0 t} \mathcal{V}_{\nu_j}(q_{\nu_j},s_j)\, \, \, ,
\end{eqnarray}
where ${\psi}_{\nu_j}$ is the field operator of the $j$ particle, $\bar{q}_{\nu_e} \, \, \left(\bar{q}_{\nu_j}\right)$ is the spatial component of the tetra-momentum of the electron-neutrino ($j$-mass-eigenstate)\cite{cahn:2013}, $\mathcal{V}_{\nu_j}$ is the anti-neutrino Dirac spinor and $q_{\nu_j}^0=\sqrt{|\bar{q}_{\nu_e}|^2+m_j^2}$ is the $j$-neutrino energy, and $m_j$ its mass.

In order to obtain the final expression for the decay-rate we write the currents in Eq.(\ref{hamil}) in terms of the particle and antiparticle fields, make all needed summations on Lorentz and spin-indexes, and integrate on their momenta.
The result is given by the expression:

\begin{eqnarray}\label{gnos}
\Gamma_{osc}&=&\frac{V^3}{2T(2\pi)^9}|A_\beta|^2\nonumber \\
&=&\frac{2}{(2\pi)^8m_n} G^2_f |V_{ud}|^2\sum_j |U_{ej}|^2\int \frac{d^3q_e}{q_{e}^0}\int \frac{d^3q_p}{q_{p}^0} \int \frac{d^3q_{\nu_j}}{q_{\nu_j}^0} e^{-(\bar{q}_p+\bar{q}_e+\bar{q}_{\nu_j})^2\delta^2}\nonumber \\
&&{\delta^0(q_p^0+q_e^0+q_{\nu_j}^0-m_n)} \times \left[-(1-g_a^2)m_p m_n \left(q_{\nu_j} \cdot q_e \right)\right. \nonumber \\
&&\left.+(1+g_a)^2 \left(q_p \cdot q_{\nu_j}\right) \left(q_n \cdot q_e \right)\right. \nonumber \\
&&\left. +(1-g_a)^2 \left(q_p \cdot q_e\right) \left(q_n \cdot q_{\nu_j}\right)\right]\, \, \, .
\end{eqnarray}

After solving the integrals and taking the limit $\delta \rightarrow 0$ we obtain
\begin{eqnarray}
\label{rateconosc}\label{gos}
\Gamma_{osc} &=&\sum_j\frac{|U_{ej}|^2 G^2_f |V_{ud}|^2}{(2\pi)^3 m_n} \int_{(m_e+m_{\nu_j})^2}^{(m_n-m_p)^2} dx \sqrt{1-2\frac{\mu_j}{x}+\frac{\xi_j}{x^2}}\sqrt{(M-x)^2-4m_p^2 m_n^2}\nonumber\\
&&\hskip 4cm \times \left\{(1+g_a^2) \left[\frac{1}{6}(M-x) x \left(1-2\frac{\mu_j}{x}+\frac{\xi_j}{x^2}\right) \right. \right. \nonumber \\
&&\left.\left. +\frac{2}{3} \left(\frac{M}{2}(M-x)-\frac{(M-x)^2}{4}-m_p^2 m_n^2\right) \left(1+\frac{\mu_j}{x}-2\frac{\xi_j}{x^2}\right)\right]\right. \nonumber \\
&&\left.-(1-g_a^2)m_n m_p(x-\mu_j)\right\} \, \, \, .
\end{eqnarray}
In the previous expression $M=m_n^2+m_p^2$, $\mu_j=m_e^2+m_{\nu_j}^2$, $\xi_j=m_e^2-m_{\nu_j}^2$, $x=M-2q_p^0m_n$, and
$g_a=1.2695 \pm 0.0058$ is the axial-vector coupling constant \cite{yao:2006}.
The oscillation parameters for active-neutrino mixing were taken from Refs. \cite{tamborra:2012,cahn:2013}: $\sin^2(2\theta_{13})=0.09$ and $\Delta m_{13}^2=2\times 10^{-3} \, {\rm eV}^2$. We assume a normal-mass-hierarchy for the ordering of the masses $m_j$ of the neutrino-mass eigenstates\cite{meregaglia:2016}. The parameters for active-sterile neutrino oscillations $\theta_{14}$ and $\Delta m_{14}^2$ are allowed to vary and the mixing matrix is written as

\begin{eqnarray}
U&=& \left(
\begin{array}{ccc}
c_{14} & 0 & s_{14} \\
0 & 1 & 0\\
-s_{14} & 0 & c_{14}
\end{array}
\right) \left(
\begin{array}{ccc}
c_{13} & s_{13} & 0 \\
-s_{13} & c_{13} & 0\\
 0 & 0 & 1 \end{array}
\right)
=\left(
\begin{array}{ccc}
c_{13}c_{14} & s_{13}c_{14} & s_{14} \\
-s_{13} &c_{13} & 0 \\
-c_{13}s_{14} & -s_{13}s_{14} & c_{14}
\end{array}
\right) \, .
\end{eqnarray}
where we have used the notation $c_{ij}=\cos \theta_{ij}$ and $s_{ij}=\sin \theta_{ij}$.\\

Since we are interested in the effect of active-sterile neutrino oscillations upon the abundances of heavy nuclei produced by the r-process, we have calculated the differential decay rates of the nuclei involved in the process and included them in the evolution equations needed to compute the nuclear abundances. The dimensionless factor
\begin{eqnarray}\label{eqf}
f&=&\frac{\Gamma_{osc}}{\Gamma_{no-osc}}\, \, \, ,
\end{eqnarray}
where $\Gamma_{no-osc}$ is the standard beta decay rate, measures the ratio between the calculated decay rates, with and without including neutrino mixing.

\section{Results}
\label{resultados}

For the calculation of the heavy nuclear abundances we use the r-java 2.0 open-use code, which performs calculations of rapid neutron capture processes in a core-collapse SN environment \cite{charignon:2011,kostka:2014-nuc,kostka:2014-astr}. The reactions included in the formalism are $\alpha$-emission \cite{lang:1980}, $\alpha$-capture, beta-decay \cite{moller:1995,moller:2003},   beta decay of delayed neutrons \cite{moller:2003}, neutron-emission, neutron-capture \cite{goriely:2008}, photo-dissociation \cite{goriely:2008} and fission. The fission channel is included in two different ways: i)through a cut-off in which the elements with A or Z that exceed the imposed limit are fragmented into two pieces, ii)in a more realistic treatment that includes spontaneous fission \cite{kodama:1975}, neutron-induced fission and beta-delayed fission \cite{panov:2005}. The weight of the fission barriers and the rates of neutron-induced fission were extracted from \cite{goriely:2008,goriely:2009}. For the realistic treatment of fission the code takes into account  three channels of mass-fragmentation and the evaporation of neutrons, which is explicitly handled for each fission event. The probability that the fission event follows a particular channel is parametrized by the relative strength and by the depth of the energy potential \cite{benlliure:1997,jurado:2014}. These parameters were calculated by adjusting observations of the mass distribution of fission fragments for nuclei between $^{232}$Th and $^{248}$Cm \cite{chadwick:2006}.

\begin{figure}[!h]
\epsfig{figure=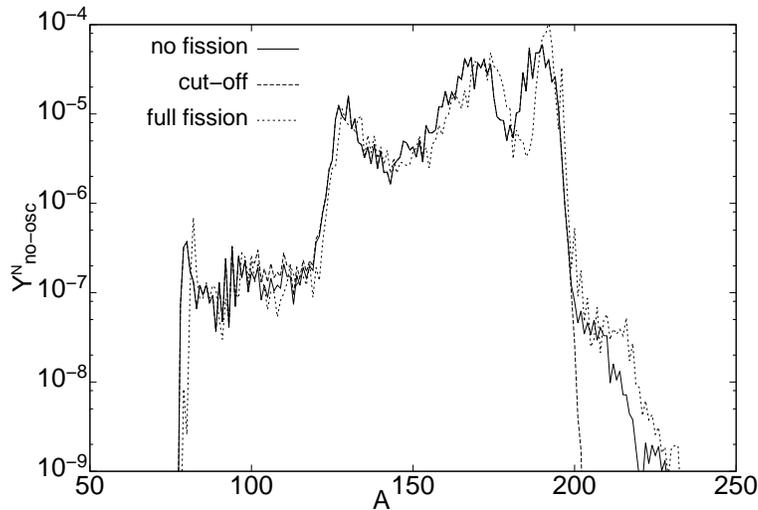, width=200 pt, angle=-90}
\caption{Nuclear abundances without considering neutrino oscillations $\left(Y_{no-osc}^N\right)$ as function of the mass number for the different treatments of fission. Solid line: no fission; dashed line: fission channel with a cut-off factor at $A=206$; dotted line: full (unrestricted) fission.}
\label{fisiones-sin-osc}
\end{figure}

We have solved the full system of equations until neutron freeze-out occurs, that is when the value of the abundance of neutrons relative to protons reaches unity, and for the high-entropy wind scenario \cite{kostka:2014-nuc}. To characterize the wind, we have chosen the initial temperature $T_0=3 \times 10^9 \, {\rm K}$, the density profile $\rho=\frac{{\rho}_0}{ (1+t/(2\tau))^2}$ where $\rho_0=10^{11} \, {\rm g/cm}^3$ is the initial density and $\tau=0.1 {\rm s}$ \cite{kostka:2014-astr,charignon:2011,meyer:1992,meyer:1997}. The initial electron fraction considered is $Y_e=0.3$, the wind speed of expansion $V_{\rm exp}=7500 \, {\rm km/s}$ and the initial wind radius $R_0=390 \, {\rm km}$ \cite{tamborra:2012,wu:2014}.
Figure \ref{fisiones-sin-osc} shows the results
for the calculated abundances, without including neutrino oscillations, and for different parametrizations of the fission channel.
The abundances for nuclei around $A=100$ for the no-fission case are smaller than those obtained by allowing fission. The results  corresponding to the full fission case agree with the ones reported in Ref. \cite{petermann:2008}, obtained by using  SMOKER \cite{panov:2005} and ALBA \cite{gaimard:1991}.

\begin{figure}[!h]
\epsfig{figure=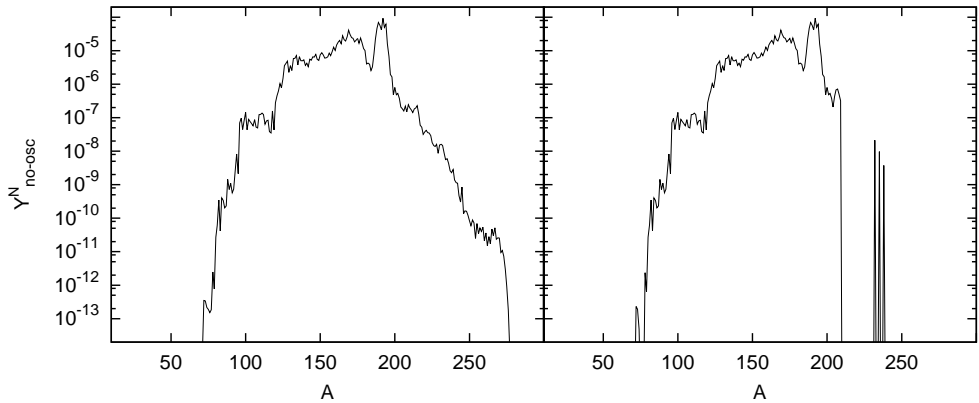, width=150 pt, angle=-90}
\caption{Nuclear abundances $\left(Y_{no-osc}^N\right)$ as a function of the mass number, considering
only beta decay and neutron capture, without including neutrino oscillations. Left column: $t_{sta}=0 \, {\rm s}$;
right column: $t_{sta}= 4.354 \times 10^{17} \, {\rm s}$.}
\label{decay-sin-osc}
\end{figure}

In  Figure \ref{decay-sin-osc} we show the results for nuclear
abundances calculated without including neutrino oscillations $\left(Y_{no-osc}^N\right)$ \cite{moller:1995,moller:2003}, for
two different values of the time needed to reach stability after the r-process, $t_{sta}$. An arbitrary large
value of the stability time, as the one used in the calculations, implies that the nuclei are allowed to decay, but
in spite of this we found nuclear species with masses up to A=$238$ (e.g: $^{232}$Th ($N=142$ and $Z=90$), $^{235}$U
and $^{238}$U ($N=143$ and $N=146$, $Z=92$), etc.

Neutrino oscillations are turned on by replacing in the code the
conventional expressions for the decay rate $\Gamma$ of
Eq.(\ref{gnos}) by the decay rates in presence of neutrino
oscillations, $\Gamma_{osc}$ of Eq.(\ref{gos}). Table 1 shows the
changes affecting the beta decay rates due to neutrino oscillations
and mixing, as expressed by the factor $f$ of Eq.(\ref{eqf}). For
all cases shown in the table, we have used the values $\Delta
m^2_{14} = 1 \, {\rm eV}^2 $, and for the active sector the
parameters $\sin^2(2\theta_{13 }) = $ 0.09 and $ \Delta m_{13}^2 = 2
\times 10^{-3} {\rm eV}^2 $. It is seen that the factor depends
strongly on the value of the mixing angle $\Theta_{14}$ between
active and sterile neutrino species. Naturally, realistic values of
the ratio $f$ are limited by the known estimates of beta decay
transitions and their renormalization factors in nuclei. What we are
showing in this table is the dependence of $f$ with the mixing
angle. The deviations from the values obtained with structureless
neutrinos are of the order of few percents, for larger values of
$\Theta_{14}$.

\begin{table}[pt]\label{rategf}
\tbl{Calculated partial decay rate, Eq. ($\ref{gos}$), and the ratio
$f$ of Eq. ($\ref{gnos}$), as a function of the mixing angle
$\Theta_{14}$. } {\begin{tabular}{@{}ccc@{}} \toprule $\Theta_{14}$&
$\Gamma_{osc}$& ratio f \\ \colrule
$ 0 $       & 0.00105 & 1.00000  \\
$ \pi / 100 $ & 0.00113 & 0.99901 \\
$ \pi / 50 $  & 0.00112 & 0.99606 \\
$ \pi / 20 $ & 0.00110 & 0.97553  \\
\botrule
\end{tabular}}
\end{table}

We have used different oscillation parameters to describe the mixing
between active neutrinos and a sterile neutrino, and took two values of the
stability time $t_{sta}$. The code was used with the initial distribution of
mass-fractions corresponding to an entropy of $2.36$ (in units of 100 times the Boltzmann
 constant) \cite{farouqi:2010}, this value is consistent with the late cooling-time of supernovae
 and with the generation of the neutrino driven wind. The results of the calculations are presented next.

\subsection{Results including neutron-capture and beta-decay in the calculation of nuclear abundances.}
\label{sec:3int}
\begin{figure}[!h]
\epsfig{figure=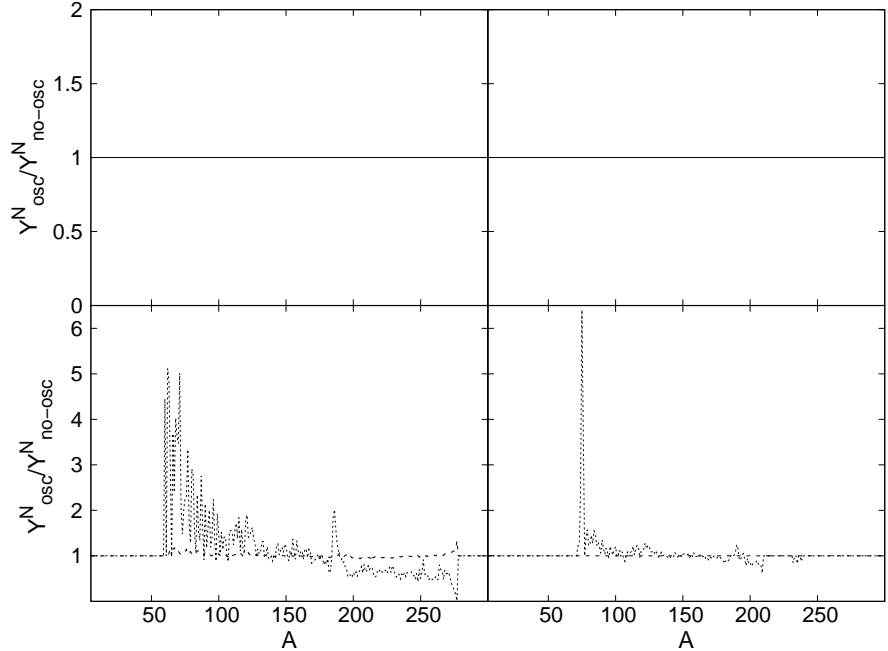, width=400 pt, angle=0}
\caption{Ratio between the nuclear abundances calculated with and without neutrino
oscillations $\left(Y_{osc}^N/Y_{no-osc}^N\right)$ as function of the mass number, considering
only beta decay and neutron capture. Left column: $t_{sta}=0 \, {\rm s}$; right
column: $t_{sta}= 4.354 \times 10^{17} \, {\rm s}$. Top panel: active-active oscillations. Bottom
panel: active-sterile oscillations; dotted line: $\theta_{14}=\pi/5$; dashed line: $\theta_{14}=\pi/10$. For all
the cases with neutrino oscillations $\Delta m^2_{14}=1 \, {\rm eV}^2$, $\sin^2 2\theta_{13}=0.09$ and $\Delta m_{13}^2=2\times 10^{-3} \, {\rm eV}^2$.}
\label{decay-con}
\end{figure}

Figure \ref{decay-con} shows the rate between the abundances calculated with and without including
neutrino oscillations, for two different values of the stability time $t_{sta}$. In the caption
to the figure we indicate the values of
$\Delta m^2_{13}$,
$\Delta m^2_{14}$, $\sin^2 2\theta_{13}$ and $\theta_{14}$.
The other parameters for neutrino-oscillations are taken from Ref.\cite{tamborra:2012,cahn:2013}.
It is seen from the results displayed in Figure \ref{decay-con}
that, when active neutrino-oscillations are included in the
treatment, both the beta-decay rate and abundance of nuclei remain
invariant with respect to the case without oscillations (top panel
of  Figure \ref{decay-con}). Instead, for the case in which the
oscillation with a sterile neutrino is turned-on the abundances
increase for nuclear masses up to A=100 and decrease for masses
larger than A$\approx$180 respect to the standard case (bottom
panel). For larger values of the mixing angle $\theta_{14}$ this
effect becomes larger. The abundances that are most affected are
those corresponding to the nuclei in the region with values of A
between 70 and 125, in particular $^{80}\rm{Se}$, $^{114}\rm{Cd}$,
$^{115}\rm{In}$ $^{121}\rm{Sb}$ , $^{123}\rm{Sb}$ , $^{126}
\rm{Te}$, even for small mixing angles $\theta_{14}$.

Since the time to stability was fixed to zero, meaning that the
nuclei reach stability once they are produced, a larger abundance of
heavy elements with A$>$200 is found (left inset of Figure
\ref{decay-con}). The effect produced by active-sterile
neutrino-mixing is smaller for larger values of the time needed to
reach stability after the r-process, as one can observed from the
curves shown in the right-hand inset of Figure \ref{decay-con}.

\subsection{Results taken a complete set of reaction in the equations.}\label{sec:allint}

We study the problem considering all the reactions which determine
the nuclear abundances, that is $\alpha$-decay \cite{lang:1980}, $\alpha$-capture, beta-decay \cite{moller:1995,moller:2003}, beta-decay of
 delayed neutrons \cite{moller:2003}, neutron-emission, neutron-capture \cite{goriely:2008}, photo-dissociation \cite{goriely:2008} and fission.

\begin{figure}[!h]
\epsfig{figure=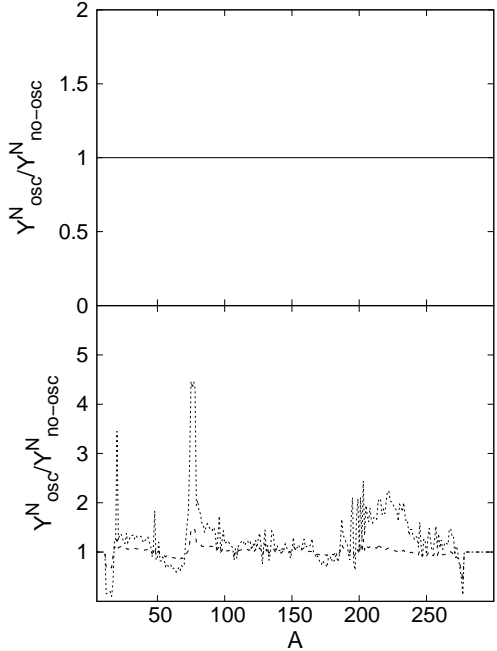, width=200 pt, angle=0}
\caption{Ratio between the abundances calculated with and without neutrino
oscillations $\left(Y_{osc}^N/Y_{no-osc}^N\right)$, as function of the mass number, considering
the realistic treatment of fission and for $t_{sta}=0 \, {\rm s}$. Top panel: active-active oscillations. Bottom
panel: active-sterile oscillations; dotted line: $\theta_{14}=\pi/5$; dashed line: $\theta_{14}=\pi/10$. For all the
cases with neutrino oscillations $\Delta m^2_{14}=1 \, {\rm eV}^2$, $\sin^2(2\theta_{13})=0.09$ and $\Delta m_{13}^2=2\times 10^{-3} \, {\rm eV}^2$.}
\label{full-t0}
\end{figure}

In Figure \ref{full-t0} we show the ratio between
the nuclear abundances with and without including neutrino-mixing, and
for $t_{sta}=0$. Again, the oscillations between active flavors do not
generate changes with respect to the standard case, that is without neutrino
oscillations. The inclusion of active-sterile neutrino mixing increases the
abundances for A$>$75 with respect to the standard case.  This effect is
different than the one obtained in the previous section. A peak indicating an
overproduction around $^{80}\rm{Se}$ appears in all cases studied, the peak
intensifies when the mixing angle is larger. The parameters adopted for the
mixing with the sterile neutrino sector are given in the caption to the figure.

\begin{figure}[!h]
\epsfig{figure=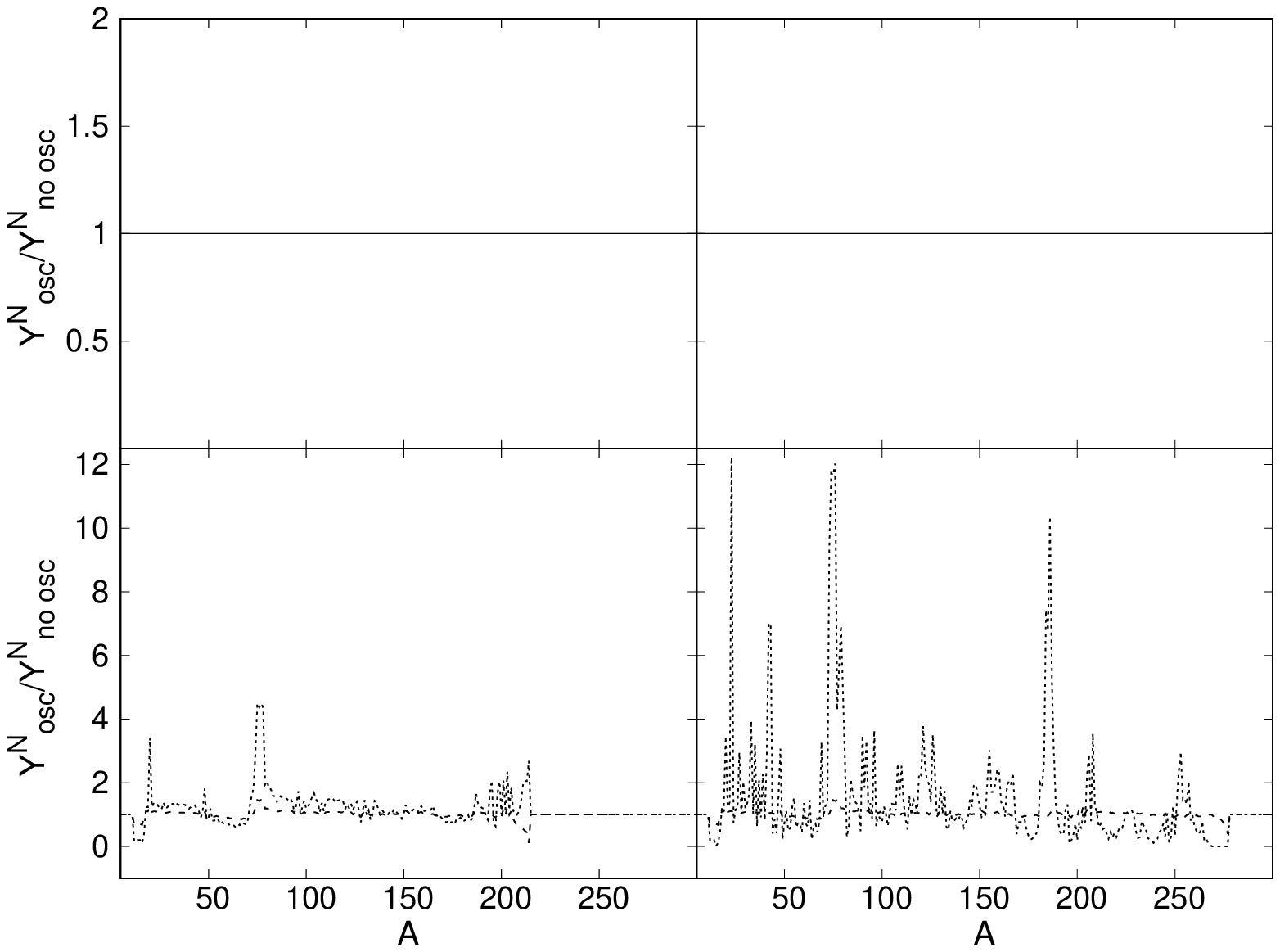, width=400 pt, angle=0}
\caption{Ratio between the
abundances calculated with and without neutrino
oscillations $\left(Y_{osc}^N/Y_{no-osc}^N\right)$ as
function of the mass number. Left panel: using a cut-off, for fission, at
the mass $A=206$; right panel: no-fission case. Top panel: active-active oscillations. Bottom
panel: active-sterile oscillations; dotted line: $\theta_{14}=\pi/5$; dashed
line: $\theta_{14}=\pi/10$. For all the cases with neutrino oscillations $\Delta m^2_{14}=1 \, {\rm eV}^2$, $\sin^2(2\theta_{13})=0.09$
and $\Delta m_{13}^2=2\times 10^{-3} \, {\rm eV}^2$. In all cases $t_{sta}=0 \, {\rm s}$.}
\label{fisiones}
\end{figure}

The results of the calculation of the abundances of
heavy-mass nuclei, excluding fission, are presented in Figure \ref{fisiones}. Also in
this case the inclusion of active-sterile neutrino-mixing has an effect upon the predicted abundances.

\section{Conclusions}
\label{conclusiones}

In this work we have calculated beta-decay-rates in presence of neutrino oscillations and used them
to calculate differential decay-rates entering in rapid neutron-capture-processes leading to
the production of heavy mass nuclei in core-collapse SN.

We found that the mixing between active neutrino flavors does not affect the decay rate or the
abundances produced by the rapid process.
In contrast, we have found that the beta-decay rate
decreases when the neutrino-mixing with a sterile specie is taken into account. The impact
on the calculated neutron-decay-rate depends strongly on the mixing angle $\theta_{14}$, and
weakly on the mass-square difference $\Delta m^2_{14}$. Accordingly, we have renormalized the
decay rates of heavy nuclei and used them to calculate nuclear abundances in the context of the r-process.

For the case in which only the main reactions are considered the change in beta decay rates, due
to active-sterile neutrino mixing, generates changes in the abundances. In particular, for A $<$ 150, the
nuclear abundances increase while for A$>$ 150  they decrease, both cases with respect to the results obtained
in standard calculations which omit neutrino mixing. The changes in the predicted nuclear abundances are larger
for larger values of the mixing angle $\theta_{14}$.

The abundances of the elements $^{80}\rm{Se}$, $^{114}\rm{Cd}$, $^{115}\rm{In}$ $^{121}\rm{Sb}$ , $^{123}\rm{Sb}$ and $^{126} \rm{Te}$, are
 the most sensitive against the inclusion of active-sterile mixing, even for small mixing angles $\theta_{14}$.

Also, we have studied the behavior of abundances for different
values of the time needed to reach stability. If we do not allow the
nuclei to decay, that is by taking $t_{sta}=0 \, {\rm s}$, the code
predicts a large abundance of heavy-mass nuclei with A$>$ 200.
Letting the nuclei decay towards stability, that is for a finite
value of $t_{sta}$, the abundance of nuclei with A$>$200
concentrates around $A=232-238$.

Finally, we have performed the analysis including all the reactions
in the equations solved by the code. It is found that the predicted abundances
change when the active-sterile neutrino-mixing is included in the calculations.

We may conclude by saying that the inclusion
of massive active and sterile neutrinos and their mixing seems
to be relevant to estimate the viability of the r-process and thus to predict the final abundances of nuclear masses.


\section*{{\bf Acknowledgment}}
This work was supported by a grant (PIP-616) of the National Research Council of Argentina (CONICET), and by a research-grant
(PICT No. 140492 ) of the National Agency for the Promotion of Science and Technology (ANPCYT) of Argentina. O. C. and M. E. M. are members of the Scientific Research Career of the CONICET, M. M. S. is a Ph.D fellow of the CONICET.


\begin{thebibliography}{99}

\bibitem{wu:2015} M.-R. Wu, G. Martinez-Pinedo, and Y.-Z. Qian, in \emph{13th International Symposium on Origin of Matter and Evolution of the Galaxies (OMEG2015) Beijing, China, June 24-27, 2015} (2015), 1512.030630.
\bibitem{qian:1993} Y. Z. Qian, G. M. Fuller, G. J. Mathews, R. W. Mayle, J. R. Wilson, and S. E. Woosley, Phys. Rev. Lett. {\bf 71}, 1965 (1993).
\bibitem{woosley:1994} S. E. Woosley,  J. R. Wilson, G. J. Mathews, R. D. Hoffman, and B. S. Meyer, ApJ {\bf 433}, 229 (1994).
\bibitem{civitarese:2014} O. Civitarese, M. E. Mosquera, and M. M. S\'aez, International Journal of Modern Physics E {\bf 23}, 1450080 (2014).
\bibitem{mosquera:2011} M. E. Mosquera, and O. Civitarese, Phys. Rev. C {\bf 84}, 065803 (2011).
\bibitem{mosquera:2015} M. E. Mosquera, and O. Civitarese, Journal of Cosmology and Astro-Particle Physics {\bf 2015}, 038 (2015).
\bibitem{saez:2018} M.~M.~Saez, O.~Civitarese and M.~E.~Mosquera, Int.\ J.\ Mod.\ Phys.\ D {\bf 27}, no. 12, 1850116 (2018).
\bibitem{aguilar:2001} A. Aguilar-Arevalo {\it et al.} (LSND), Phys. Rev. D {\bf 64}, 112007 (2001).
\bibitem{fukuda:1998} Y.~Fukuda {\it et al.} (Super-Kamiokande), Phys.\ Rev.\ Lett.\  {\bf 81}, 1562 (1998).
\bibitem{poon:2001} A.~W.~P.~Poon {\it et al.} (SNO), AIP Conf.\ Proc.\  {\bf 610}, 218 (2002).
\bibitem{aguilar:2007} A. A. Aguilar-Arevalo et al. (MiniBoNE), Phys. Rev. Lett. {\bf 98}, 231801 (2007).
\bibitem{aguilar:2013}  A. A. Aguilar-Arevalo et al. (MiniBoNE), Phys. Rev. Lett. {\bf 110}, 161801 (2013).
\bibitem{palomares:2009} C. Palomares (Double Chooz), PoS {\bf EPS-HEP2009}, 275 (2009).
\bibitem{abdurashitov:2009} J.~N.~Abdurashitov {\it et al.} (SAGE), Phys.\ Rev.\ C {\bf 80}, 015807 (2009).
\bibitem{detwiler:2003} J. A. Detwiler (KamLAND), eConf {\bf C0307282}, TW04 (2003).
\bibitem{arpesella:2008} C.~Arpesella {\it et al.} (Borexino), Phys.\ Rev.\ Lett.\  {\bf 101}, 091302 (2008).
\bibitem{eitel:2000} K. {Eitel}, New Journal of Physics {\bf 2}, 1 (2000).
\bibitem{athana:1995} C. Athanassopoulos et al., Phys. Rev. Lett. {\bf 75}, 2650 (1995).
\bibitem{athanassopoulos:1996} C. Athanassopoulos et al. (LSND), Phys. Rev. Lett. {\bf 77}, 3082 (1996).
\bibitem{aguilar:2018} A.~A.~Aguilar-Arevalo {\it et al.} (MiniBooNE), Phys.\ Rev.\ Lett.\  {\bf 121}, no. 22, 221801 (2018).
\bibitem{giunti:2011}  C.~Giunti and M.~Laveder, Phys.\ Rev.\ D {\bf 84}, 073008 (2011).
\bibitem{himmel:2015} A.~Himmel, Physics Procedia {\bf 61}, 612 (2015), 13th International Conference on Topics in Astroparticle and Underground Physics, TAUP 2013.
\bibitem{boyarsky:2009} A. Boyarsky, O. Ruchayskiy and M. Shaposhnikov, Ann. Rev. Nucl. Part. Sci. {\bf 59}, 191 (2009).
\bibitem{mohapatra:2004} R. N. Mohapatra and P. B. Pal, \emph{Massive Neutrinos in Physics and Astrophysics}, Lecture Notes in Physics Series (World Scientific, 2004). ISBN 9789812380715.
\bibitem{raffelt:1987} G. G. Raffelt, and D. Seckel, Phys. Rev. Lett. {\bf 60}, 1793 (1988).
\bibitem{molinari:2003} A. Molinari, L. Riccati and W. M. Alberico, \emph{From Nuclei and Their Constituents to Stars} International School of Physics Enrico Fermi (IOS Press, 2003), ISBN 9781614990093.
\bibitem{balasi:2015} K. G. Balasi, K. Langanke, and G. Martinez-Pinedo, Prog. Part. Nucl. Phys. {\bf 85}, 33 (2015).
\bibitem{fetter:2003} J. Fetter, G. C. McLaughlin, A. B. Balantekin and G. M. Fuller, Astropart. Phys. {\bf 18}, 433 (2003).
\bibitem{balantekin:2004} A. Balantekin and H. Yuksel, New J. Phys. {\bf 7}, 51 (2005).
\bibitem{tamborra:2012} I. Tamborra, G. G. Raffelt, and D. V Semikoz, JCAP {\bf 1}, 13 (2012).
\bibitem{janka:2012} H.-T. Janka, Ann. Rev. Nucl. Part. Sci. {\bf 62}, 407 (2012).
\bibitem{pastor:2002} S. Pastor, G. G. Raffelt, and D. V. Semikoz, Phys. Rev. D {\bf 65}, 053011 (2002).
\bibitem{qian:2003} Y.-Z. Qian, Prog. Part. Nucl. Phys. {\bf 50}, 153 (2003).
\bibitem{anders:1989} E.~Anders and N.~Grevesse, Geochim.\ Cosmochim.\ Acta {\bf 53}, 197 (1989).
\bibitem{duan:2010} H.~Duan, A.~Friedland, G.~C.~McLaughlin and R.~Surman, J.\ Phys.\ G {\bf 38}, 035201 (2011).
\bibitem{kasen:2017} D. {Kasen} {\it et al.}, Nature {\bf 551}, 7678 (2017)
\bibitem{cowan:2019} J.~J.~Cowan, C.~Sneden, J.~E.~Lawler, A.~Aprahamian, M.~Wiescher, K.~Langanke, G.~Martínez-Pinedo and F.~K.~Thielemann,
 arXiv:1901.01410 [astro-ph.HE].
\bibitem{curtis:2018} S.~Curtis, K.~Ebinger, C.~Fröhlich, M.~Hempel, A.~Perego, M.~Liebendörfer and F.~K.~Thielemann,
  Astrophys.\ J.\  {\bf 870}, no. 1, 2 (2019).
\bibitem{clayton:1968} D. D. Clayton, \emph{Principles of stellar evolution and nucleosynthesis: with a new preface} (University of Chicago Press, 1968), ISBN 9780226109527.
\bibitem{marshak:1969} R. E. Marshak, Riaduzzin and C. P. Ryan, \emph{Theory of weak interactions in particle physics} (Wiley-Interscience; New York, 1969).
\bibitem{blin-stoyle:1973} R.J. Blin-Stoyle, \emph{Fundamental Interactions and the Nucleus [By] R.J. Blin-Stoyle} (North-Holland Publishing Company; New York: American Elsevier Publishing Company, 1973).
\bibitem{yao:2006} W. M. Yao et al., Journal of Physics G {\bf 33}, 1 (2006).
\bibitem{ivanov:2008} A.~N.~Ivanov, R.~Reda and P.~Kienle, (2008), 0801.2121.
\bibitem{cahn:2013} R. N. {Cahn} D. A. Dwyer, S. J. Freedman, W. C. Haxton, R. W. Kadel, Yu. G. Kolomensky, K. B. Luk, P. McDonald, G. D. Orebi Gann, and A. W. P. Poon, in \emph{Proceedings, 2013 Community Summer Study on the Future of U.S. Particle Physics: Snowmass on the Mississippi (CSS2013): Minneapolis, MN, USA, July 29-August 6, 2013} (2013), 1307.5487.
\bibitem{meregaglia:2016} A. Meregaglia. (The Double Chooz Collaboration), Nuovo Cimento Geophysics Space Physics C, {\bf 38}, 123  (2016).
\bibitem{charignon:2011} C.~Charignon, M.~Kostka, N.~Koning, P.~Jaikumar and R.~Ouyed, Astron.\ Astrophys.\  {\bf 531}, A79 (2011).
\bibitem{kostka:2014-nuc}  M.~Kostka, N.~Koning, Z.~Shand, R.~Ouyed and P.~Jaikumar, Astron.\ Astrophys.\  {\bf 568}, A97 (2014).
\bibitem{kostka:2014-astr} M.~Kostka, N.~Koning, Z.~Shand, R.~Ouyed and P.~Jaikumar (2014), arXiv:1402.3824 [astro-ph.IM].
\bibitem{lang:1980} K. R. Lang, \emph{Astrophysical Formulae. A Compendium for the Physicist and Astrophysicist.}
(Springer-Verlag Berlin Heidelberg New York. Also Springer Study Edition, 1980).
\bibitem{moller:1995} P. {M\"oller}, J. R. {Nix}, W. D. {Myers}, and W. J. {Swiatecki}, Atomic Data and Nuclear Data Tables {\bf 59}, 185 (1995).
\bibitem{moller:2003} P. {M\"oller},  B. {Pfeiffer}, and K. L. {Kratz}, {Phys. Rev. C} {\bf 67}, 055802 (2003).
\bibitem{goriely:2008} S.~Goriely, S.~Hilaire and A.~J.~Koning, Astron.\ Astrophys.\  {\bf 487}, 767 (2008).
\bibitem{kodama:1975} T. {Kodama} and K. {Takahashi}, Nuclear Physics A {\bf 239}, 489 (1975).
\bibitem{panov:2005} I. V. Panov, E. Kolbe,  B. {Pfeiffer}, T. Rauscher, K. L. {Kratz}, and F. K. Thielemann, Nucl. Phys. A {\bf 747}, 633 (2005).
\bibitem{goriely:2009} S. Goriely, S. Hilaire, A. J. {Koning}, M. {Sin}, and R. {Capote}, Phys. Rev. C {\bf 79}, 024612 (2009).
\bibitem{benlliure:1997} J.~Benlliure, K.~H.~Schmidt, A.~Grewe, M.~de Jong and S.~Zhdanov, Nucl.\ Phys.\ A {\bf 628}, 458 (1998).
\bibitem{jurado:2014} B.~Jurado and K.~H.~Schmidt, J.\ Phys.\ G {\bf 42}, no. 5, 055101 (2015).
\bibitem{chadwick:2006} M. B. {Chadwick} et al., Nuclear Data Sheets {\bf 107}, 2931 (2006).
\bibitem{meyer:1992} B. S. Meyer,  G. J. {Mathews},  W. M.  {Howard}, S. E.{Woosley} and R. D. {Hoffman}, ApJ {\bf 399}, 656 (1992).
\bibitem{meyer:1997} B. S. Meyer and J. S. Brown, The Astrophysical Journal Supplement Series {\bf 112}, 199 (1997).
\bibitem{wu:2014} M.~R.~Wu, T.~Fischer, L.~Huther, G.~Martínez-Pinedo and Y.~Z.~Qian, Phys.\ Rev.\ D {\bf 89}, no. 6, 061303 (2014).
\bibitem{petermann:2008} I.~Petermann, K.~Langanke, G.~Martinez-Pinedo, P.~v.~Neumann-Cosel, F.~Nowacki and A.~Richter, Phys.\ Rev.\ C {\bf 81}, 014308 (2010).
\bibitem{gaimard:1991} J.~J.~Gaimard and K.~H.~Schmidt, Nucl.\ Phys.\ A {\bf 531}, 709 (1991).
\bibitem{farouqi:2010} K. {Farouqi}, K. L. {Kratz}, B. {Pfeiffer}, T. {Rauscher}, F. K. {Thielemann}, and J. W. {Truran}, ApJ {\bf 712}, 1359 (2010).
\end{thebibliography}
\end{document}